\documentclass[11pt]{article}

\newcommand{\beq}{\begin{equation}}
\newcommand{\eeq}{\end{equation}}
\newcommand{\p}{\partial}
\newcommand{\bfe}{\mbox{\bf e}}
\newcommand{\bfu}{\mbox{\bf u}}
\newcommand{\bfp}{\mbox{\bf p}}

\begin{document}

\vspace*{1.in}

\begin{center}

{\Large {\bf Farewell to General Relativity}} \\

\vspace{.5in}

{\large {\bf Kenneth Dalton}} \\

\vspace{.3in}

    email: kxdalton@yahoo.com

\vspace{1.5in}

{\bf Abstract}

\end{center}

\vspace{.25in}

The kinematical successes of general relativity are legendary: the 
perihelion precession, the gravitational red-shift, the bending of light.
However, at the level of dynamics, relativity is faced with insurmountable
difficulties.  It has failed to define the energy, momentum, and stress of
the gravitational field.  Moreover, it offers no expression of energy-momentum
transfer to or from the gravitational field.  These are symptoms of a far
graver malady: general relativity violates the principle of energy
conservation.

\clearpage

In general relativity, the equation of planetary motion 
is derived by means of the geometric variation 

\beq
  \delta \int ds = \delta \int \sqrt{g_{\mu\nu} dx^\mu dx^\nu} = 0
\eeq
This yields the geodesic equation

\beq
   \frac{du^\mu}{ds} 
      + \Gamma^{\mu}_{\nu\lambda} u^\nu u^\lambda = 0
\eeq
The equation of motion can be recast in terms of kinematics, 
by introducing a vector basis $ {\bfe}_\mu $ and the velocity four-vector
$ {\bf u} = {\bfe}_\mu u^\mu . $ The infinitesimal change of the basis
is expressed in terms of connection coefficients
\beq
      d {\bfe}_\mu = {\bfe}_\lambda \, \Gamma^\lambda_{\mu\nu} \, dx^\nu
\eeq
which enables the calculation

\begin{eqnarray}
   \frac{d {\bf u}}{ds} 
   & = &  \left\{ {\bfe}_\mu \frac{du^\mu}{ds}
       + \frac{d {\bfe}_\mu}{ds} u^\mu \right\}  \nonumber \\
   & = &  {\bfe}_\mu \left\{ \frac{du^\mu}{ds}
       + \Gamma^\mu_{\nu\lambda} u^\nu u^\lambda \right\}
\end{eqnarray}
This shows that $ \, d {\bf u}/ds = 0 \,$ along any geodesic path.

How do these formulae relate to the observed planetary motion?  
Expanding $\, {\bf u} = {\bfe}_0 u^0 + {\bfe}_i u^i\, $  we have

\beq
       \frac{d({\bfe}_0 u^0)}{ds} + \frac{d({\bfe}_i u^i)}{ds} = 0
\eeq
This shows that, during geodesic motion, the rate of change of
three-velocity $ \, {\bfe}_i u^i \, $ is equal and opposite to that 
of speed $ \, {\bfe}_0 u^0 \, $. The rates are determined by (2) and (3)

\beq
     \frac{d({\bfe}_0 u^0)}{ds} = - \frac{d({\bfe}_i u^i)}{ds} 
     = - {\bfe}_0 \Gamma^0_{i\nu} u^i u^\nu
       + {\bfe}_i \Gamma^i_{0\nu} u^0 u^\nu
\eeq
In flat spherical coordinates, the right-hand side of this equation is zero: 
both speed and velocity are constant.  In Schwarzschild coordinates, 
the right-hand side is not zero: the speed and velocity continually
change.  Thus, according to the geodesic hypothesis, the changes
which we observe in a planet's speed and velocity are due 
to the curved geometry of space-time.    

The success of the geodesic formula was one of the great triumphs of 
twentieth-century physics.  Yet, we know that a planet possesses dynamical
properties of energy and momentum.  What is taking place at this, 
the dynamical level?  The energy-momentum vector of a planet is

\begin{eqnarray}
                \bfp & = & m \bfu  \nonumber \\
                     & = & {\bfe}_\mu p^\mu = {\bfe}_0 p^0 +{\bfe}_i p^i 
\end{eqnarray}
The first term is the (rest + kinetic) energy $ \, {\bfe}_0 p^0 \,$ while the 
second term is the momentum $ \, {\bfe}_i p^i \,$.  During geodesic motion, 
the energy-momentum of the planet is conserved 

\beq    
     \frac{d \bf p}{ds} = m \frac{d \bf u}{ds}
    = m{\bf e}_\mu \left\{\frac{du^\mu}{ds} 
      + \Gamma^\mu_{\nu\lambda}u^\nu u^\lambda \right\} = 0
\eeq
Therefore, the rates of change of energy 
and momentum are equal and opposite

\beq
     \frac{d({\bfe}_0 p^0)}{ds} = - \frac{d({\bfe}_i p^i)}{ds}
\eeq
Neither the planet's energy nor its momentum is conserved; rather, one
continually transforms into the other.  During orbital motion, the 
non-conservation of linear momentum is to be expected.  What surprises is
that energy conservation is violated.  The energy principle forces us to 
abandon the geodesic hypothesis of planetary motion.

The treatment of light-rays is similar to that of particle motion, in that 
the bending of light and the gravitational red-shift are determined by the 
kinematics of curved space-time.  The red-shift can be expressed in terms
of the null four-vector

\beq
     {\bf k} = {\bf e}_0 k^0 + {\bf e}_i k^i
\eeq
where $ \, {\bfe}_0 k^0 \, $ is the frequency of light, and $ \, {\bfe}_i k^i \, $
is its wave vector.  Along any light ray $ \, d {\bf k}/d \lambda = 0 \, $
and we obtain 

\beq
 \frac{d({\bfe}_0 k^0)}{d \lambda} = - \frac{d({\bfe}_i k^i)}{d \lambda} 
     = - {\bfe}_0 \Gamma^0_{i\nu} k^i u^\nu
       + {\bfe}_i \Gamma^i_{0\nu} k^0 u^\nu
\eeq
Therefore, in the presence of space-time curvature, the frequency and 
wavelength will vary from point to point along the light ray.  This is the
gravitational red-shift.
\newpage
The energy-momentum vector of a light complex is given by the quantum
formula

\beq
     {\bf p} = \hbar {\bf k}
\eeq
Once again, energy-momentum is conserved $\, d {\bf p}/d \lambda = 0 \,$ and

\beq
     \frac{d({\bfe}_0 p^0)}{d \lambda} = - \frac{d({\bfe}_i p^i)}{d \lambda}     
\eeq
Thus, as frequency and wavelength change, the energy and momentum 
transform into one another; neither is conserved.

\begin{center}
----------------------------
\end{center}

The above examples illustrate the violation of energy conservation during
geodesic motion.  We will now make use of the field equations to show that
this problem is intrinsic to the theory, and stems from the fact that the
gravitational field is incapable of exchanging energy-momentum with any
physical field.  The field equations are given by

\beq
     R^{\mu\nu} - \frac{1}{2} g^{\mu\nu} R = - \kappa\, T^{\mu\nu}
\eeq
where $ T^{\mu\nu} $ is the stress-energy-momentum tensor of matter and
electromagnetism.  The covariant divergence of the left-hand side is 
identically zero, therefore

\beq
   T^{\mu\nu}_{;\nu} =
         \frac{1}{\sqrt{-g}}
         \frac{\p \sqrt{-g}\,T^{\mu \nu}}{\p x^\nu}
         + \Gamma^\mu_{\nu\lambda} T^{\nu\lambda}
         = 0
\eeq
Let us investigate this equation, by way of three examples from classical
physics.

Consider a free electromagnetic field, with the energy tensor

\beq
     T^{\mu\nu}_{\it e-m} = F^\mu_{\;\alpha} F^{\alpha\nu}
       + \frac{1}{4} g^{\mu\nu} F_{\alpha\beta} F^{\alpha\beta}
\eeq
A lengthy but straightforward calculation yields

\beq
 T^{\mu\nu}_{;\nu \,\, {\it e-m}} 
   = \frac{1}{\sqrt{-g}}\frac{\p\sqrt{-g}\, F^{\alpha\nu}}{\p x^\nu}
       F^\mu_{\;\alpha}
    + \frac{1}{2} g^{\mu\nu} F^{\alpha\beta}
     \left\{\frac{\p F_{\beta\nu}}{\p x^\alpha}
     + \frac{\p F_{\nu\alpha}}{\p x^\beta}
     + \frac{\p F_{\alpha\beta}}{\p x^\nu} \right\}
\eeq
We note that the term $ \Gamma^\mu_{\nu\lambda} T^{\nu\lambda} $
must be included, in order to obtain this covariant expression.
Maxwell's equations for charge-free space are 

\beq
   \frac{1}{\sqrt{-g}}\frac{\p\sqrt{-g}\, F^{\alpha\nu}}{\p x^\nu} = 0
\eeq

\beq
       \frac{\p F_{\beta\nu}}{\p x^\alpha} +
       \frac{\p F_{\nu\alpha}}{\p x^\beta} +
       \frac{\p F_{\alpha\beta}}{\p x^\nu} 
         = 0
\eeq
and we obtain

\beq
     T^{\mu\nu}_{;\nu \;\; {\it e-m}} = 0
\eeq
This result is especially significant, because it shows that there is no
mechanism whatsoever for the exchange of energy-momentum between the 
electromagnetic and gravitational fields.  The energy-momentum of 
electromagnetism alone is conserved.

Secondly, consider the matter tensor

\beq
     T^{\mu\nu}_{\it m} = \rho u^\mu u^\nu
\eeq
A simple calculation yields the covariant expression

\beq
   T^{\mu\nu}_{;\nu \;\; {\it m}} 
    = \rho u^\nu \frac{\p u^\mu}{\p x^\nu}
   +  u^\mu \frac{1}{\sqrt{-g}} \frac{\p \sqrt{-g} \, \rho u^\nu}{\p x^\nu}
   + \Gamma^\mu_{\nu\lambda} \,\rho u^\nu u^\lambda
\eeq       
The second term is zero, if rest mass is conserved, leaving
 
\beq
T^{\mu\nu}_{;\nu \;\; {\it m}} 
  = \rho \left\{u^\nu \frac{\p u^\mu}{\p x^\nu}
    + \Gamma^\mu_{\nu\lambda} \, u^\nu u^\lambda \right\}
\eeq
The hydrodynamical form of the geodesic equation then gives

\beq
    T^{\mu\nu}_{;\nu \;\; {\it m}} = 0
\eeq
The energy-momentum of matter alone is conserved.

Finally, consider the case of charged matter together with
electromagnetism

\beq
     T^{\mu\nu} = T^{\mu\nu}_{\it m}+ T^{\mu\nu}_{\it e-m}
\eeq
Here, coupling occurs via Maxwell's equation

\beq
  \frac{1}{\sqrt{-g}}\frac{\p\sqrt{-g} F^{\alpha\nu}}{\p x^\nu} 
          = - j^\alpha     
\eeq
and we obtain

\beq
 T^{\mu\nu}_{;\nu} =
     \rho \left\{u^\nu \frac{\p u^\mu}{\p x^\nu}
     + \Gamma^\mu_{\nu\lambda} \, u^\nu u^\lambda \right\} 
     - j^\alpha F^\mu_{\;\alpha} 
            = 0
\eeq
The Lorentz force is covariant and describes the exchange of 
energy-momen\-tum between matter and the electromagnetic field.

These examples show that whether space-time is curved or not, i.e.,
whether a gravitational field exists or not, the energy-momentum of matter
and electromagnetism is conserved.  It follows that any change wrought by 
curvature---in speed, velocity, frequency, and wavelength---will violate
the principle of energy conservation.  A gravitational exchange term is
needed in order to account for the changes in energy and momentum. 
The theory of relativity neither provides such a term nor defines the 
energy, momentum, and stress of the gravitational field.  
If we adhere to the energy principle, then general relativity 
cannot be the answer to the question of gravitation.

\end{document}